\documentclass[a4paper,12pt,table]{article}
\usepackage{amssymb,amsmath,latexsym,color,amsthm,authblk,mathpazo,palatino,bbding,setspace,datetime,breakcites,hyphenat,microtype,diagbox,tikz,xcolor,tkz-graph,subcaption,graphicx,titlesec,accents}
\usepackage[round,authoryear]{natbib}
\usepackage{hyperref,theoremref}
\usepackage[titletoc,title]{appendix}
\usepackage{physics}
\usepackage[shortlabels]{enumitem}
\usepackage{float}
\usepackage{amssymb}
\restylefloat{table}
\usepackage[margin=1.25in]{geometry}

\everymath{\displaystyle}

\definecolor{armygreen}{rgb}{0.29, 0.33, 0.13}
\definecolor{auburn}{rgb}{0.43, 0.21, 0.1}
\definecolor{burgundy}{rgb}{0.5, 0.0, 0.13}
\definecolor{medium red}{rgb}{.490,.298,.337}
\definecolor{dark red}{rgb}{.235,.141,.161}

\hypersetup{
	colorlinks = true,
	linkcolor = {burgundy},
	citecolor = {burgundy}, 		
	linkbordercolor = {white},
}

\let\OLDthebibliography\thebibliography
\renewcommand\thebibliography[1]{
	\OLDthebibliography{#1}
	\setlength{\parskip}{0pt}
	\setlength{\itemsep}{0pt plus 0.1ex}
}

\DeclareFontFamily{U}{mathx}{\hyphenchar\font45}
\DeclareFontShape{U}{mathx}{m}{n}{<-> mathx10}{}
\DeclareSymbolFont{mathx}{U}{mathx}{m}{n}
\DeclareMathAccent{\widebar}{0}{mathx}{"73}

\newcommand*{\TitleFont}{%
	\usefont{\encodingdefault}{\rmdefault}{b}{n}%
	\fontsize{16}{20}%
	\selectfont}

    \titleformat{\paragraph}
{\normalfont\scshape\normalsize}{\theparagraph}{1em}{}
\titlespacing*{\paragraph}
{0pt}{3.25ex plus 1ex minus .2ex}{1.5ex plus .2ex}

\titleformat{\section}[block]{\normalfont\scshape\normalsize\filcenter}{\thesection .}{1em}{}
\titleformat{\subsection}{\normalfont\scshape\normalsize}{\thesubsection}{1em}{}
\titleformat{\subsubsection}{\normalfont\scshape\normalsize}{\thesubsubsection}{1em}{}

\newtheorem{theorem}{Theorem}[section]

\theoremstyle{definition}
\newtheorem{definition}{Definition}[section]

\theoremstyle{remark}

\theoremstyle{observation}
\newtheorem{observation}{\textsc{Observation}}[section]

\newdateformat{monthyeardate}{\monthname[\THEMONTH], \THEYEAR}

\title{\TitleFont \textsc{On the existence of EFX allocations for goods}\thanks{\noindent The authors would like to thank Florian Brandl, Debasis Mishra, Benny Moldovanu and Soumyarup Sadhukhan for their helpful discussions and suggestions. The authors would also like to thank the audience of BGSE Workshop, University of Bonn for their helpful comments. Ujjwal Kumar acknowledges that his work was supported by the Deutsche Forschungsgemeinschaft (DFG, German Research Foundation) under Germany's Excellence Strategy - GZ 2047/1, Projekt-ID 390685813.}}

\author[1]{Ujjwal Kumar}
\author[2]{Souvik Roy\footnote{Corresponding author: souvik.2004@gmail.com}}
\affil[1]{Hausdorff Center for Mathematics and Institute for Microeconomics\\ University of Bonn}
\affil[2]{Statistical Sciences Division, Indian Statistical Institute, Kolkata}

\date{\monthyeardate\today}

\begin{document}
	
	\maketitle
	\vspace*{-10mm}
	
\begin{abstract}
We consider a set $E$ of $m$ indivisible goods and a set $N$ consisting of $n \geq 2$ agents. The paper shows that if two agents have \textit{arbitrary} set monotonic valuation functions and the remaining agents have size monotonic valuation functions, then EFX  allocations always exist.
\end{abstract}

    \section{Introduction}
	Our goal is to divide a finite set of indivisible goods among $n$ agents in a ``fair" manner. The fairness notion that we consider is envy based, that is, agents compare their allocated bundle of goods with the bundle of goods allocated to other agents. An allocation is said to be Envy Free ($EF$) if no agent envies any other agent. EF allocations might not exist, for example, in the case of one good and two agents, no allocation is EF. To overcome this difficulty, \citet{caragiannis2019unreasonable} introduced the notion of envy-freeness upto any good (EFX) allocations. An allocation is said to be $EFX$ if no agent envies any other agent following the removal of \textit{any} single good from the other agent's allocated bundle.
	
	We consider a set $E$ of $m$ indivisible goods and a set $N$ consisting of  $n \geq 2$ agents.
	A valuation function maps every bundle (subset of goods) to a real number specifying the valuation of the bundle. A valuation function is set monotonic if valuation increases as more goods are added to a bundle and it is size monotonic if bundles with higher number of goods have higher valuations. Clearly, size monotonicity implies set monotonicity. The question whether EFX allocations always exist when agents have set monotonic valuation functions is open till date and is considered one of the most enigmatic open problem in fair division.

	\citet{plaut2020almost} show that EFX allocations always exist when all agents have identical set monotonic valuation functions and also when $n=2$. \citet{akrami2023efx} show that EFX allocations always exist when there are 3 agents and atleast one of them have MMS-feasible valuation function (see Definition \ref{MMS} for details) while the remaining agents have arbitrary set monotonic valuation function. \citet{mahara2023existence} shows the that EFX allocations always exist for any number of agents with atmost two distinct additive set monotonic valuation functions.

	\subsection{Contribution of the Paper}
	
	Our main result shows that if two agents have \textit{any} set monotonic valuation functions and the remaining agents have size monotonic valuation functions, then EFX allocations always exist. The full force of set monotonicity and size monotonicity is not needed for this theorem, in Theorem \ref{relax_2} we show that our main theorem holds under much weaker technical conditions. We do not present the weaker version here.
	
	Despite significant efforts from researchers, the existence of EFX allocations has been shown for very few instances. The problem is still open for $n\geq 3$ when all agents have arbitrary set monotonic valuation functions. To identify an EFX allocation, generally the starting point is an allocation (say $X$, possibly specific or arbitrary) and if $X$ is not EFX, then the subsequent steps involve moving among the space of allocations till we find an EFX allocation, which makes the problem notoriously difficult. What we do to prove our main result is that we start with fixing an agent (say Agent 1) and our starting point is a carefully designed allocation $X^1=(X_1,\ldots,X_n)$ such that $X_1$ is the best bundle of size 1 according to Agent 1's valuation function and the remaining $X_i's$ are such that either $X^1$ is EFX or Agent 1 envies some agent.\footnote{We assume that all agents except Agent 1 and Agent 2 have size monotonic valuation functions.} It is worth noting that such allocation $X^1$ always exists because all but two agents have size monotonic valuation functions. The reason for this lies behind the fact that the structure of the allocation $X^1$ is such that the size of the bundle $X_2$ (say $r$) is atmost one less than the size of the bundle $X_j$ where $j \in \{3,\ldots,n\}$ and $X_2$ is the best bundle of size $r$ chosen from $E \setminus X_1$ according to Agent 2's valuation function. All but two agents having size monotonic valuation functions ensure that there is no envy in terms of (or "with respect to") EFX between any two agents $i,j \in \{2,\ldots,n\}$. In the subsequent steps, we move in the space of allocations such that at each step either we get an EFX allocation or we get an allocation (say $X^k$) such that $i$ does not envy $j$ for all $i,j \in \{2,\ldots,n\}$ and $|X^{k-1}_1|<|X^k_1|$. Once again, all but two agents having size monotonic valuation functions is a crucial requirement to get hold of such allocations $X^k$. Therefore, at each step either we get an EFX allocation (can be $X^k$) or we consider the allocation $X^k$ where Agent 1 envies some other agent. This process must terminate as $|X^k_1|$ cannot increase beyond a certain \textit{threshold} because of the nature of our carefully contructed allocation $X^k$. Therefore, at the final step (where the process terminates), we get hold of an EFX allocation.  
    
    To the best of our knowledge, our paper is the first one to prove the existence result for any number of agents where every agent can possibly have distinct set monotonic valuation functions. Recently, \citet{hv2024efx} show the existence result for any number of agents when each agent is assigned a valuation function from a prespecified (but arbitrary) set of valuation functions containing \textit{three} distinct \textit{additive} valuation function. Also, our proof techniques yield a new proof (for the existence result) when $n=2$, which is significantly different than the proof in \citet{plaut2020almost} and proved to be quite insightful for proving our general result.
	
	Two important open problems in this area are  the existence of EFX for $n \geq 3$ when agents have arbitrary set monotonic valuation functions, and for $n \geq 4$ agents having  additive set monotonic valuation functions.

	\section{Model}
	Let $E$ be a set of $m$ indivisible goods and $N=\{1,\ldots,n\}$ be the set of agents where $n \geq 2$.
	A valuation function is a mapping $v:2^E \to \mathbb{R}$ that specifies the valuation of each subset of $E$.\footnote{We denote by $2^E$ the set of all subsets of $E$.}
	
	A valuation function $v$ is set monotonic if for all $X, X' \in  2^E$ with $X \subseteq X' $, we have $v(X) \leq v(X')$. Let  $1\leq k<l\leq m$. A valuation function $v$ is called $\big \langle k,l \big \rangle$- set monotonic if for all $X \subseteq Y \subseteq E$ with $k \leq |X|<|Y|\leq l$, we have  $v(X) \leq v(Y)$. Notice that a valuation function $v$ is set monotonic if $v$ is $\big \langle 1,m \big \rangle$- set monotonic. A valuation function $v$ is called $\big \langle k,l \big \rangle$-size monotonic if for all $X \neq  Y \subseteq E$ with $k \leq |X|<|Y|\leq l$, we have  $v(X) \leq v(Y)$.

	Each agent $i \in N$ has a set monotonic valuation function $v_i$. A collection of set monotonic valuation functions $v_N:= (v_i)_{i \in N}$ is called a set monotonic valuation profile. A collection $X_N:= (X_i)_{i \in N}$ of subsets of $E$ is called an allocation if $\{X_1,\ldots,X_n\}$ forms a partition of $E$.\footnote{A collection of (possibly empty) subsets $\{X_1,\ldots,X_n\}$ of $X$ is a partition of $X$ if $X_i \cap X_j=\emptyset$ for every pair $i,j \in N$ and $\bigcup_{i \in N}X_i=E$.}
	
	Given a valuation function $v$ and two subsets $X$ and $Y$ of $E$, we say \textit{$X$ is envy-free from $Y$ up to any one item at $v$}, and write $X E_{v} Y$, if  $v(X) \geq v(Y \setminus \{y\})$ for all $y \in Y$.
	
	\begin{definition}
		An allocation $X_N$ is said to be envy-free upto any good (EFX) at a valuation profile $v_N$ if $X_i E_{v_i} X_j$ for all $i,j \in N$.
	\end{definition}
	
	\begin{observation} \label{m>n}
		An EFX allocation exists at every set monotonic valuation profile whenever $m \leq n$. In particular, any allocation $X_N$ where $|X_i|\leq 1$ for each $i \in N$ is EFX at every set monotonic valuation profile. Note that one does not need the full force of set monotonicity for $X_N$ to be EFX; $X_N$ continues to be EFX as long as $v(\{x\}) \geq v(\emptyset)$ for each $x \in E$.
	\end{observation}
	
	In view of Observation \ref{m>n}, we assume $m>n$ for the rest of the paper. 
	
	A valuation function $\hat{v}$ is strict if $\hat{v}(X) \neq \hat{v}(Y)$ for all $X \neq Y$. A valuation profile $\hat{v}_N$ is called strict if each valuation function $\hat{v}_i$ is strict. To minimize confusion, we sometimes refer to an arbitrary valuation function (where the valuations could be strict or weak) as a weak valuation function. 
	
	\begin{observation} \label{strict valuations}
		Let $\bar{v}_N$ be a (weak) valuation profile and $\hat{v}_N$ be a \textit{strict}  valuation profile such that for each $i \in N$ and for all distinct $X,Y \in 2^E$, $\bar{v}_i(X)> \bar{v}_i(Y)$ implies $\hat{v}_i(X)> \hat{v}_i(Y)$. Suppose $X_N$ is an EFX allocation at $\hat{v}_N$. Then, $X_N$ is also EFX at $\bar{v}_N$. 
	\end{observation}
	To see how Observation \ref{strict valuations} follows, assume for contradiction that $\bar{v}_i(X_i) < \bar{v}_i(X_j \setminus \{x_j\})$ for some $i,j \in N$ and some $x_j \in X_j$. Then, by our assumption $\hat{v}_i(X_i) < \hat{v}_i(X_j \setminus \{x_j\})$, implying that $X_N$ is not EFX at $\hat{v}_N$, a contradiction.

	\section{Results}
	
	\citet{plaut2020almost} show that EFX allocations always exist when all agents have identical set monotonic valuation functions and also when $n=2$.

	\begin{definition} \label{MMS}
		A set monotonic valuation function $v$ is called MMS-feasible if for all $X \subseteq E$ and every two partitions $\{X_1,X_2 \}$ and $\{X'_1,X'_2\}$ of $X$, we have $\max \{v(X_1),v(X_2)\} \geq \min \{v(X'_1),v(X'_2)\}$.
	\end{definition}
	
	\begin{theorem} \label{akrami} [\citet{akrami2023efx}] An EFX allocation exists at a set monotonic valuation profile $v_N$ if $n=3$ and $|\{i \in N :  v_i \mbox{ is MMS-feasible } \}| \geq n-2$.
	\end{theorem}

	Since $m>n$, we have $m=n \times  \big \lfloor m/n \big \rfloor + m\big|_n$ where $\big \lfloor m/n \big \rfloor$ is the greatest integer that is less than or equal to $m/n$ and $m\big|_n := m \; (\mbox{mod }n)$.  The numbers $\big \lfloor m/n \big \rfloor$ and $m\big|_n$ play important roles in our results. 
	
	\begin{theorem} \label{n-1 size monotonic}
		Let $v_N$ be a set monotonic valuation profile.
		\begin{enumerate}[(i)]
			\item Suppose $0\leq m\big|_n \leq 1$. Then, there exists an EFX allocation at $v_N$ if $n-1$ agents have $\big  \langle \big \lfloor m/n \big \rfloor-1,\big \lfloor m/n \big \rfloor \big \rangle$-size monotonic valuation functions.
			\item Suppose $1<m\big|_n \leq n-1$. Then, there exists an EFX allocation at $v_N$ if $m\big|_n$ agents have $\big \langle \big \lfloor m/n \big \rfloor,\big \lfloor m/n \big \rfloor+1 \big \rangle$-size monotonic valuation functions, and a different set of $n-1-m\big|_n$ agents have $\big \langle \big \lfloor m/n \big \rfloor-1,\big \lfloor m/n \big \rfloor \big \rangle$-size monotonic valuation functions.
			
		\end{enumerate}
	\end{theorem}
	
	The proof of this theorem is relegated to Appendix \ref{A}.

	Since $m>n$, it follows that $m-1>n-1$. Hence, we have $m-1=(n-1) \times  \big \lfloor (m-1)/(n-1) \big \rfloor + (m-1)\big|_{(n-1)}$. 
	
	\begin{theorem} \label{n-2 size monotonic}
		Let $v_N$ be a set monotonic valuation profile.
		\begin{enumerate}[(i)]
			\item Suppose $0\leq m-1\big|_{n-1} \leq 1$. Then, there exists an EFX allocation at $v_N$ if $n-2$ agents have $\big \langle 1,\big \lfloor m-1/n-1 \big \rfloor \big \rangle$-size monotonic valuation functions.
			\item Suppose $1<m-1\big|_{n-1}\leq n-2$. Then, there exists an EFX allocation at $v_N$ if $m-1\big|_{n-1}$ agents have $\big \langle 1, \big \lfloor m-1/n-1 \big \rfloor+1 \big \rangle$-size monotonic valuation functions, and a different set of $n-2-m-1\big|_{n-1}$ agents have $\big \langle 1,\big \lfloor m-1/n-1 \big \rfloor \big \rangle$-size monotonic valuation functions.
		\end{enumerate}
		
		The proof of this theorem is relegated to Appendix \ref{B}.

	\end{theorem}

	It is worth emphasizing that neither size monotonicity implies the MMS-feasibility nor does the MMS-feasibility imply size monotonicity. To see this, let $E=\{a,b,c,d\}$ be a set of 4 goods. 
	
	First, we show that size monotonicity does not imply MMS-feasibility by giving an example of a size monotonic valuation function that is not MMS-feasible. For any $\ell \in \{1,2\}$, consider any $\big \langle \ell, \ell+1 \big \rangle$-size monotonic valuation function $v$ such that $v(\{a,b\})>v(\{c,d\})>v(\{b,c\})>v(\{a,d\})$. Note that such $v$ exists because size monotonicity does not put any restriction on bundles of goods having the same cardinality (in this case, bundles having cardinality 2). $v$ is \textit{not} MMS-feasible because for two partitions $\{\{a,b\},\{c,d\}\}$ and $\{\{b,c\},\{a,d\}\}$ of $E$, $\max\{v(\{b,c\}),v(\{a,d\})\}<\min\{v(\{a,b\}),v(\{c,d\})\}$. Hence, size monotonicity does not imply MMS-feasibility. 
	
	Next, we show that MMS-feasibility does not imply size monotonicity by giving an example of a MMS-feasibile valuation function that is not size monotonic. A valuation function $v$ is additive if $v(S)=\sum_{s \in S}v(s)$ for all $S \subseteq E$. \citet{akrami2023efx} show that every additive valuation function is also MMS-feasible. Consider any additive (hence, MMS-feasible) valuation function $v$ such that $v(\{a\})>v(\{b,c,d\})$. Since we assume valuation functions to be set monotonic, it must be the case that $v(S) \leq v(\{b,c,d\})$ for all $S \subseteq \{b,c,d\}$ and $v(S) \geq v(\{a\})$ for all $S \subseteq E$ such that $a \in S$. We claim that for each $\ell \in \{1,2\}$, $v$ is not $\big \langle \ell, \ell+1 \big \rangle$-size monotonic. This is because when $\ell=1$, $v(\{a\})>v(\{b,c\})$ implies that $v$ is not $\big \langle 1, 2 \big \rangle$-size monotonic. Similarly, when $\ell=2$, $v(\{a,b\})>v(\{b,c,d\})$ implies that $v$ is not $\big \langle 2, 3 \big \rangle$-size monotonic. Therefore, for each $\ell \in \{1,2\}$. $v$ is not $\big \langle \ell, \ell+1 \big \rangle$-size monotonic. Hence, MMS-feasibility does not imply size monotonicity.
	
	In particular, Theorem \ref{n-2 size monotonic} for three agents is independent of the result in \citet{akrami2023efx} (Theorem \ref{akrami}).
	
	\section{Discussions}
	
We observe that the full force of set monotonicity is not required for the valuation profiles in Theorems~\ref{n-1 size monotonic} and~\ref{n-2 size monotonic}. In this section, we relax the assumption of set monotonicity. The proof techniques used in the original theorems remain valid even under these weaker conditions. However, we choose not to include these relaxed versions in the main text, as they may be less appealing or accessible to the broader readership.

Theorems~\ref{relax_1} and~\ref{relax_2} are generalized versions of Theorems~\ref{n-1 size monotonic} and~\ref{n-2 size monotonic}, obtained by relaxing the requirement of set monotonicity.
While Theorem~\ref{n-1 size monotonic} assumes that \emph{all} agents have set-monotonic valuation functions, Theorem~\ref{relax_1} allows for more flexibility. Specifically, it permits some agents (see the theorem statement for precise details) to have $\langle \lfloor m/n \rfloor, \lfloor m/n \rfloor+1 \rangle$-set monotonic or $\langle \lfloor m/n \rfloor-1, \lfloor m/n \rfloor \rangle$-set monotonic valuation functions.
Similarly, Theorem~\ref{relax_2} relaxes the assumption in Theorem~\ref{n-2 size monotonic} by allowing some agents to have $\langle \lfloor (m-1)/(n-1) \rfloor, \lfloor (m-1)/(n-1) \rfloor+1 \rangle$-set monotonic or $\langle 1, \lfloor (m-1)/(n-1) \rfloor \rangle$-set monotonic valuation functions, instead of requiring full set monotonicity across all agents.
As noted earlier, $\langle k, l \rangle$-size monotonicity implies $\langle k, l \rangle$-set monotonicity for any $1 \leq k < l \leq m$.
	
	\begin{theorem} \label{relax_1}
		Let $v_N$ be an arbitrary valuation profile.
		\begin{enumerate}[(i)]
			\item Suppose $0\leq m\big|_n\leq 1$. Then, there exists an EFX allocation at $v_N$ if $m\big|_n$ agents have $\big \langle \big \lfloor m/n \big \rfloor-1,\big \lfloor m/n \big \rfloor \big \rangle$-size monotonic  and $\big \langle \big \lfloor m/n \big \rfloor,\big \lfloor m/n \big \rfloor+1 \big \rangle$-set monotonic valuation functions, a different set of $n-1-m\big|_n$ agents have $\big \langle \big \lfloor m/n \big \rfloor-1,\big \lfloor m/n \big \rfloor \big \rangle$-size monotonic valuation functions, and the remaining agent has $\big \langle \big \lfloor m/n \big \rfloor-1,\big \lfloor m/n \big \rfloor \big \rangle$-set monotonic valuation function.

			\item Suppose $1< m\big|_n \leq n-2$. Then, there exists an EFX allocation at $v_N$ if $m\big|_n$ agents have $\big \langle \big \lfloor m/n \big \rfloor,\big \lfloor m/n \big \rfloor+1 \big \rangle$-size monotonic  and $\big \langle \big \lfloor m/n \big \rfloor-1,\big \lfloor m/n \big \rfloor \big \rangle$-set monotonic valuation functions, a different set of $n-1-m\big|_n$ agents have $\big \langle \big \lfloor m/n \big \rfloor-1,\big \lfloor m/n \big \rfloor \big \rangle$-size monotonic valuation functions, and the remaining agent has $\big \langle \big \lfloor m/n \big \rfloor-1,\big \lfloor m/n \big \rfloor \big \rangle$-set monotonic valuation function.
			
			\item Suppose $m\big|_n= n-1$. Then, there exists an EFX allocation at $v_N$ if $n-1$ agents have $\big \langle \big \lfloor m/n \big \rfloor,\big \lfloor m/n \big \rfloor+1 \big \rangle$-size monotonic  and $\big \langle \big \lfloor m/n \big \rfloor-1,\big \lfloor m/n \big \rfloor \big \rangle$-set monotonic valuation functions.
		\end{enumerate}
	\end{theorem}

	\begin{theorem} \label{relax_2}
		Let $v_N$ be an arbitrary valuation profile.
		\begin{enumerate}[(i)]
			\item Suppose $0\leq m-1\big|_{n-1}\leq 1$. Then, there exists an EFX allocation at $v_N$ if $m-1\big|_{n-1}$ agents have $\big \langle 1,\big \lfloor m-1/n-1 \big \rfloor \big \rangle$-size monotonic  and $\big \langle \big \lfloor m-1/n-1 \big \rfloor,\big \lfloor m-1/n-1 \big \rfloor+1 \big \rangle$-set monotonic valuation functions, a different set of $n-2-m-1\big|_{n-1}$ agents have $\big \langle 1,\big \lfloor m-1/n-1 \big \rfloor \big \rangle$-size monotonic valuation functions, and the remaining two agents have $\big \langle 1,\big \lfloor m-1/n-1 \big \rfloor \big \rangle$-set monotonic valuation function.

			\item Suppose $1< m-1\big|_{n-1} \leq n-2$. Then, there exists an EFX allocation at $v_N$ if $m-1\big|_{n-1}$ agents have $\big \langle 1,\big \lfloor m-1/n-1 \big \rfloor+1 \big \rangle$-size monotonic valuation functions, a different set of $n-2-m-1\big|_{n-1}$ agents have $\big \langle 1,\big \lfloor m-1/n-1 \big \rfloor \big \rangle$-size monotonic valuation functions, and the remaining two agents have $\big \langle 1,\big \lfloor m-1/n-1 \big \rfloor \big \rangle$-set monotonic valuation function.
			
		\end{enumerate}
		
	\end{theorem}

	\begin{appendices}
		{
			\def\OldComma{,}
			\catcode`\,=13
			\def,{%
				\ifmmode%
				\OldComma\discretionary{}{}{}%
				\else%
				\OldComma%
				\fi%
			}
			
			\section{Appendix}			
			
			\subsection{Proof of Theorem \ref{n-1 size monotonic}} \label{A}
			\vspace{2mm}
			
			\noindent
			\begin{proof}
				In view of Observation \ref{strict valuations}, we prove Theorem \ref{n-1 size monotonic} when valuation profiles are strict. We give a unified proof here that works for both parts of the theorem. Let $\hat{v}_N$ be a strict set monotonic valuation profile. Also, let $r=m\big|_n$ and $\ell=\big \lfloor m/n \big \rfloor$. By the assumption of the theorem, $n-1$ agents have $\big \langle k,k+1 \big \rangle$-size monotonic valuation functions at $\hat{v}_N$ where $k \in \{\ell-1,\ell\}$ accordingly as in the two parts of the theorem. Assume without loss of generality that $\hat{v}_1$ is not necessarily $\big \langle k,k+1 \big \rangle$-size monotonic where $k \in \{\ell-1,\ell\}$ accordingly as in the two parts of the theorem. For part (ii) of the theorem, also assume without loss of generality that for each $i\in \{n-r+1,\ldots,n\}$, $v_i$ is $\big \langle \ell,\ell+1 \big \rangle$-size monotonic. 
				
				Let $\mathcal{X}_{0}= \emptyset$ and $\mathcal{X}_{i} = \cup_{j=1}^{i}X_j$ for $i \geq 1$. Define the allocation $X_N$ as follows: 
				\begin{enumerate}
					\item [(i)] for $1 \leq i \leq n-r$, $X_i=\arg\hspace{-8.8mm}\max_{S \subseteq E \setminus \mathcal{X}_{i-1} : |S|=\ell}\hat{v}_i(S)$,\footnote{This is well-defined (that is, the argmax is unique) since $\hat{v}_N$ is strict.} and
					\item[(ii)] for $n-r<i\leq n$, $X_i \subseteq E \setminus \mathcal{X}_{i-1}$ such that $|X_i|=\ell+1$.
				\end{enumerate}
				
				By the definition, $X_i\cap X_j =\emptyset$ for every $i,j \in N$, and $\sum_{i=1}^{n}|X_i|=n\ell+r=m$, which justifies that $X_N$ is indeed an allocation. We show that the allocation $X_N$ is EFX at $\hat{v}_N$. Consider an arbitrary agent $i \in N$, another agent $j \in N \setminus \{i\}$, and any good $x_j \in X_j$. We show that $\hat{v}_i(X_i) \geq {\hat{v}_i} (X_j \setminus \{x_j \})$.
				
				Suppose $i=1$. By the definition of $X_N$, $|X_1|= \ell$ and $|X_j| \in \{\ell,\ell+1\}$. Suppose $|X_j|=\ell+1$. By the definition of $X_1$, we have $\hat{v}_1(X_1) \geq \hat{v}_1(S)$ for all $S \subseteq E$ with $|S|=\ell$.\footnote{By definition, $X_1=\arg\hspace{-3.5mm}\max_{S \subseteq E : |S|=\ell}\hat{v}_1(S)$, which in other words means that $\hat{v}_1(X_1) \geq \hat{v}_1(S)$ for all $S \subseteq E$ with $|S|=\ell$.} This, together with the fact that $|X_j \setminus \{x_j\}|=\ell$, implies $\hat{v}_1(X_1) \geq {\hat{v}_1} (X_j \setminus \{x_j \})$. Now, suppose $|X_j|=\ell$. Since $|X_j|=\ell$ and $\hat{v}_1$ is strict, by the definition of $X_1$, $\hat{v}_1(X_1)> \hat{v}_1(X_j)$. By the set monotonicity (in particular, by $\big \langle \ell-1, \ell \big \rangle$-set monotonicity) of $\hat{v}_1$, we have $\hat{v}_1(X_j)> \hat{v}_1(X_j \setminus \{x_j\})$. Combining, we obtain $\hat{v}_1(X_1)>\hat{v}_1(X_j \setminus \{x_j\})$.
				
				Suppose $i \in \{2,\ldots,n-r\}$. By the definition of $X_N$, $|X_i|=\ell$. Suppose $j \in \{1,\ldots,n-r\}$. Then, by definition $|X_j|=\ell$. This, together with the fact that $\hat{v}_i$ is $\big \langle \ell-1,\ell \big \rangle$-size monotonic, we have $\hat{v}_i(X_i) > \hat{v}_i(X_j \setminus \{x_j\})$. Now, suppose  $j \in N \setminus \{1,\ldots,n-r\}$. Then, by definition, $|X_j|=\ell+1$, and hence $|X_j \setminus \{x_j\}|=\ell$. Since $1 \leq i \leq n-r$ and $j> n-r$, which in particular means that $i<j$, we have that both $X_i$ and $X_j \setminus \{x_j\}$ are subsets of  $ E \setminus \mathcal{X}_{i-1}$, where $\mathcal{X}_{i-1}$ is as defined in the definition of $X_N$. Since $X_i=\arg\hspace{-8.8mm}\max_{S \subseteq E \setminus \mathcal{X}_{i-1} : |S|=\ell}\hat{v}_i(S)$ and $|X_i|=|X_j \setminus \{x_j\}|=\ell$, by the definition of $X_i$ it follows that $\hat{v}_i(X_i) > \hat{v}_i(X_j \setminus \{x_j\})$.

				Finally, suppose $i \in N \setminus \{1,\ldots,n-r\}$. Then, by definition, $|X_i|=\ell+1$. Furthermore, by the definition of $X_N$, $|X_k| \in \{\ell,\ell+1\}$ for all $k \in N$. For part (i) of the theorem, if $r=0$, then there is nothing to prove. If $r=1$, then $i=n, j \in \{1,\ldots,n-1\}$ and $|X_j|=\ell$. Since $\hat{v}_i$ is $\big \langle \ell-1,\ell \big \rangle$-size monotonic and also a set monotonic (in particular, by $\big \langle \ell, \ell+1 \big \rangle$-set monotonic) valuation function, it follows that $\hat{v}_i(X_i) > \hat{v}_i(X_j \setminus \{x_j\})$. For part (ii) of the theorem, note that $|X_j|$ is either $\ell$ or $\ell+1$. Therefore, $|X_j \setminus \{x_j\}|$ is either $\ell-1$ or $\ell$. Since $\hat{v}_i$ is $\big \langle \ell,\ell+1 \big \rangle$-size monotonic and also a set monotonic (in particular, by $\big \langle \ell-1, \ell \big \rangle$-set monotonic) valuation function, it follows that $\hat{v}_i(X_i) > \hat{v}_i(X_j \setminus \{x_j\})$.				
				This completes the proof.
			\end{proof}

			\subsection{Proof of Theorem \ref{n-2 size monotonic}} \label{B}
			\vspace{2mm}
			
			\noindent
			\begin{proof}
				By Observation \ref{strict valuations}, it is enough to prove the existence of an EFX allocation at each \textit{strict} monotone valuation profile $\hat{v}_N$ satisfying the assumptions of Theorem \ref{n-2 size monotonic}. We give a unified proof here that works for both parts of the theorem. Let $\hat{v}_N$ be a strict set monotonic valuation profile. Also, let $r=m-1\big|_{n-1}$ and $\ell=\big \lfloor m-1/n-1 \big \rfloor$. By the assumption of the theorem, $n-2$ agents have $\big \langle 1,k+1 \big \rangle $-size monotonic valuation functions at $\hat{v}_N$ where $k \in \{\ell-1,\ell\}$ accordingly as in the two parts of the theorem. Let $M$ be the set of such $n-2$ agents. Assume without loss of generality that $\hat{v}_1$ and $\hat{v}_2$ are not necessarily $\big \langle 1,k+1 \big \rangle $-size monotonic where $k \in \{\ell-1,\ell\}$ accordingly as in the two parts of the theorem. Therefore, $N \setminus M=\{1,2\}$. For part (ii) of the theorem, also assume without loss of generality that for each $i\in \{n-r+1,\ldots,n\}$, $v_i$ is $\big \langle 1,\ell+1 \big \rangle $-size monotonic.
				
				Let $W^1= \arg\hspace{-4.5mm}\max_{S \subseteq E : |S|=1}\hat{v}_1(S)$. Define $Y^1= \arg\hspace{-7mm}\max_{S \subseteq E \setminus W^1 : |S|=\ell}\hat{v}_2(S)$. Let $\mathcal{X}^1_{0}= \emptyset$ and $\mathcal{X}^1_{i} = \cup_{j=1}^{i}X^1_j$ for $i \geq 1$. Now, we define the allocation $X^1_N$ as follows: 
				\begin{enumerate}
					\item[(i)] $X^1_1=W^1$,
					
					\item[(ii)] $X^1_2=Y^1$,
					
					\item[(iii)] For $3 \leq i \leq n-r$, $X^1_i=\arg\hspace{-8.5mm}\max_{S \subseteq E \setminus \mathcal{X}^1_{i-1} : |S|=\ell}\hat{v}_i(S)$, and
					\item [(iv)] For $n-r<i\leq n$, $X^1_i \subseteq E \setminus \mathcal{X}^1_{i-1}$ such that $|X^1_i|=\ell+1$.
				\end{enumerate}
				
				By definition, $X^1_i\cap X^1_j =\emptyset$ for every $i,j \in N$, and $\sum_{i=1}^{n}|X^1_i|=|X^1_1|+\sum_{i=2}^{n}|X^1_i|=1+(n-1)\ell+r=1+(m-1)=m$, which justifies that $X^1_N$ is indeed an allocation. Define $\max(X^1_N)=\max\{|X^1_i| : i \in N\}$. Notice that for any $i \in N \setminus \{1\}$, $\max(X^1_N)$ is either $|X^1_i|$ or $|X^1_i|+1$. Also, note that $|E|> \max(X^1_N)$ since $n \geq 2$.
				
				Consider the set of agents $\{2,\ldots,n\}$ having cardinality $n-1$. One out of the $n-1$ agents (Agent 2) can have any arbitrary monotone valuation function whereas the remaining agents have $\big \langle 1,k+1 \big \rangle $-size monotonic valuation functions where $k \in \{\ell-1,\ell\}$ accordingly as in the two parts of the theorem. Therefore, these satisfy the conditions of Theorem \ref{n-1 size monotonic} when the set of agents is $\{2,\ldots,n\}$ instead of $\{1,2,\ldots,n\}$. Hence, by the proof of Theorem \ref{n-1 size monotonic}, it follows that $X^1_i E_{\hat{v}_i} X^1_j$ for every $i,j \in \{2,\ldots,n\}$. Also, $X^1_i E_{\hat{v}_i} X^1_1$ for every $i \in \{2,\ldots,n\}$ because $|X^1_1|=1$.\footnote{Here we use the fact that empty set is the least preferred bundle for any set monotonic valuation function.} Therefore, either $X^1_N$ is EFX at $\hat{v}_N$ or $X^1_1 E_{\hat{v}_1} X^1_i$ does not hold for some $i \in \{2,\ldots,n\}$. It is worth noting that if $\ell=1$ or $\ell=2$ and $r=0$, then $X^1_N$ must be EFX at $\hat{v}_N$ because $X^1_1 E_{\hat{v}_1} X^1_i$  holds for every $i \in \{2,\ldots,n\}$ as $|X^1_i|\leq 2$ for every $i \in \{2,\ldots,n\}$ and by definition, $X^1_1$ is the highest valued singleton good set according to $\hat{v}_1$, thereby completing the proof. So, we consider only the cases where $\ell=2$ and $r>0$ or $\ell>2$ and $r\geq 0$. Therefore, using this, one can also conclude that $\max(X_N^1) \geq 3$ and $|X^1_1|< |X^1_i|$ for every $i \in \{2,\ldots,n\}$.
				
				If $X^1_N$ is EFX at $\hat{v}_N$, then the proof is complete. We now consider the other remaining possibility. Suppose that there exists $q \in \{2,\ldots,n\}$ such that $X^1_1 E_{\hat{v}_1} X^1_q$ does not hold. This implies that $\hat{v}_1(X^1_q\setminus \{x_q\}) > \hat{v}_1(X^1_1)$ for some $x_q \in X^1_q$. Since $X^1_q\setminus \{x_q\} \subsetneq X^1_q$ and $\hat{v}_1$ is a strict set monotonic valuation function (in particular $\big \langle \ell-1, \ell+1 \big \rangle$-set monotonic), we must have $\hat{v}_1(X^1_q)>\hat{v}_1(X^1_q\setminus \{x_q\}) > \hat{v}_1(X^1_1)$. Therefore, there exists $Z \subseteq E \setminus X^1_1$ with $|Z|=\max(X^1_N)-1$ such that $\hat{v}_1(Z) > \hat{v}_1(X^1_1)$. Note that such a set $Z$ always exists because $\max(X^1_N)$ can either be $|X^1_q|$ or $|X^1_q|+1$. If $|X^1_q|=\max(X^1_N)$, we can choose $Z=X^1_q\setminus \{x_q\}$. If $|X^1_q|\neq \max(X^1_N)$, then it must be the case that $|X^1_q|+1=\max(X^1_N)$. In this case, we can choose $Z=X^1_q$. Also, notice that $E \neq Z \cup X_1^1$ because $|Z\cup X^1_1|=(\max(X^1_N)-1)+1= \max(X^1_N)$ and $|E|> \max(X^1_N)$. Therefore, $E \setminus (Z\cup X^1_1) \neq \emptyset$.
				
				Define $G^1=\{S \subseteq E : |S|=2 \mbox{ and } \exists T \subseteq E \setminus S \mbox{ with } |T|=\max(X^1_N)-1 \mbox{ such that } \hat{v}_1(T)>\hat{v}_1(X^1_1)\}$. Notice that $G^1 \neq \emptyset$ because for any $g \in E \setminus \big( Z\cup X^1_1 \big), X^1_1\cup \{g\} \in G^1$ as $|X^1_1\cup \{g\}|=2$ and $Z \subseteq E \setminus \big( X^1_1\cup \{g\} \big)$ with $|Z|=\max(X^1_N)-1$ such that $\hat{v}_1(Z)>\hat{v}_1(X^1_1)$. Define $W^2=\arg\hspace{-1mm}\max_{S \in G^1}\hat{v}_1(S)$.
				Also, let $H^1=\{S \subseteq E \setminus W^2 : |S|=\max(X^1_N)-1 \mbox{ and } \hat{v}_1(S)>\hat{v}_1(X^1_1)\}$. Since, $W^2 \in G^1$  by definition, it follows that $H^1 \neq \emptyset$. Define $Y^2=\arg\max_{S \in H^1}\hat{v}_1(S)$. Let $\mathcal{X}^2_{0}= \emptyset$ and $\mathcal{X}^2_{i} = \cup_{j=1}^{i}X^2_j$ for $i \geq 1$. Now, we define the allocation $X^2_N$ as follows:
				\begin{enumerate}
					\item [(i)] $X^2_2=\arg\hspace{-5mm}\max_{S \in \{W^2,Y^2\}}\hat{v}_2(S)$,
					
					\item [(ii)]$X^2_1=\{W^2,Y^2\} \setminus X^2_2$,
					
					\item[(iii)] For $3 \leq i \leq n-r+1$\footnote{With the interpretation that for $r=0$, $n+1\equiv2$.}, $X^2_i=\hspace{10mm}\arg\hspace{-17mm}\max_{S \subseteq E \setminus \mathcal{X}^2_{i-1} : |S|=\max(X^1_N)-1}\hat{v}_i(S)$, and
					\item[(iv)] For $n-r+1< i\leq n$, $X^2_i \subseteq E \setminus \mathcal{X}^2_{i-1}$ such that $|X^2_i|=\max(X^1_N)$.
				\end{enumerate}
				
				By definition, $X^2_i\cap X^2_j =\emptyset$ for every $i,j \in N$ and $\sum_{i=1}^{n}|X^2_i|=m$ because $|X^2_1|=|X^1_1|+1$, $|X^2_{n-r+1}|=|X^1_{n-r+1}|-1$ and $|X^2_j|=|X^1_j|$ for all $j \in N \setminus \{1,n-r+1\}$. Therefore, $X^2_N$ is indeed an allocation. 
				
				Consider the set of agents $\{3,\ldots,n\}$ having cardinality $n-2$. All these agents have $\big \langle 1,k+1 \big \rangle $-size monotonic valuation functions where $k \in \{\ell-1,\ell\}$ accordingly as in the two parts of the theorem. Therefore, these satisfy the conditions of Theorem \ref{n-1 size monotonic} when the set of agents is $\{3,\ldots,n\}$ instead of $\{1,2,3,\ldots,n\}$. Hence, by the proof of Theorem \ref{n-1 size monotonic}, it follows that $X^2_i E_{\hat{v}_i} X^2_j$ for every $i,j \in \{3,\ldots,n\}$. Also, $X^2_i E_{\hat{v}_i} X^2_j$ for $i \in \{3,\ldots,n\}$ and $j \in \{1,2\}$ because for each $i \in \{3,\dots,n\}$, $\hat{v}_i$ is $\big \langle 1,k+1 \big \rangle $-size monotonic where $k \in \{\max(X^1_N)-1,\max(X^1_N)\}$ accordingly as in the two parts of the theorem. Therefore, $X^2_i E_{\hat{v}_i} X^2_j$ for every $i \in \{3,\ldots,n\}$ and $j \in \{1,\ldots,n\}$. Also, by definition, $X^2_2 E_{\hat{v}_2} X^2_1$. We distinguish two cases: 
				
				\textbf{Case A:} Suppose $X^2_2=W^2$. Then by the definition of $Y^2$ we have $X^2_1 E_{\hat{v}_1} X^2_i$ for every $i \in \{2,\ldots,n\}$. Therefore, either $X^2_N$ is EFX at $\hat{v}_N$ or $X^2_2 E_{\hat{v}_2} X^2_i$ does not hold for some $i \in \{3,\ldots,n\}$. If $X^2_N$ is EFX at $\hat{v}_N$, then the proof is complete. Suppose $X^2_N$ is \textit{not} EFX at $\hat{v}_N$. Then $X^2_2 E_{\hat{v}_2} X^2_i$ does not hold for some $i \in \{3,\ldots,n\}$. Let $\bar{\mathcal{X}}^2_{0}= \emptyset$ and $\bar{\mathcal{X}}^2_{i} = \cup_{j=1}^{i}\bar{X}^2_j$ for $i \geq 1$. Define the allocation $\bar{X}^2_N$ as follows:
				\begin{enumerate}
					\item $\bar{X}^2_1=W^2$,
					
					\item $\bar{X}^2_2=\hspace{10mm}\arg\hspace{-16mm}\max_{S \subseteq E \setminus W^2 : |S|=\max(X^1_N)-1}\hat{v}_2(S)$,
					
					\item For $3 \leq i \leq n-r+1$\footnote{With the interpretation that for $r=0$, $n+1\equiv2$.}, $\bar{X}^2_i=\hspace{10mm}\arg\hspace{-17mm}\max_{S \subseteq E \setminus \bar{\mathcal{X}}^2_{i-1} : |S|=\max(X^1_N)-1}\hat{v}_i(S)$, and
					\item For $n-r+1<i\leq n$, $\bar{X}^2_i \subseteq E \setminus \bar{\mathcal{X}}^2_{i-1}$ such that $|\bar{X}^2_i|=\max(X^1_N)$.
				\end{enumerate}
				
				By definition, $\bar{X}^2_i\cap \bar{X}^2_j =\emptyset$ for every $i,j \in N$ and $\sum_{i=1}^{n}|\bar{X}^2_i|=\sum_{i=1}^{n}|X^2_i|=m$. Therefore, $\bar{X}^2_N$ is indeed an allocation.
				
				Consider the set of agents $\{2,\ldots,n\}$ having cardinality $n-1$. One out of the $n-1$ agents (Agent 2) can have any arbitrary monotone valuation function whereas the remaining agents have $\big \langle 1,k+1 \big \rangle $-size monotonic valuation functions where $k \in \{\ell-1,\ell\}$ accordingly as in the two parts of the theorem. Therefore, these satisfy the conditions of Theorem \ref{n-1 size monotonic} when the set of agents is $\{2,\ldots,n\}$ instead of $\{1,2,\ldots,n\}$. Hence, by the proof of Theorem \ref{n-1 size monotonic}, it follows that $\bar{X}^2_i E_{\hat{v}_i} \bar{X}^2_j$ for every $i,j \in \{2,\ldots,n\}$. Also, $\bar{X}^2_2 E_{\hat{v}_2} \bar{X}^2_1$ holds because of the definition of $\bar{X}^2_2$ and the fact that $X^2_2 E_{\hat{v}_2} X^2_i$ does not hold for some $i \in \{3,\ldots,n\}$. To see this, observe that since $X^2_2 E_{\hat{v}_2} X^2_i$ does not hold for some $i \in \{3,\ldots,n\}$, it must be the case that there exists $U \subseteq E \setminus W^2$ with $|U|=\max(X^1_N)-1$ such that $\hat{v}_2(U)>\hat{v}_2(W^2)$. Therefore, by the definition of $\bar{X}^2_2$, it follows that $\bar{X}^2_2 E_{\hat{v}_2} \bar{X}^2_1$. In addition, $\bar{X}^2_i E_{\hat{v}_i} \bar{X}^2_1$ for every $i \in \{3,\ldots,n\}$ because $|\bar{X}^2_1|=2\leq \max(X^1_N)-1$ and for each $i \in \{3,\dots,n\}$, $\hat{v}_i$ is $\big \langle 1,k+1 \big \rangle $-size monotonic where $k \in \{\max(X^1_N)-1,\max(X^1_N)\}$ accordingly as in the two parts of the theorem. Therefore, we have an allocation $\bar{X}^2_N$ such that $\bar{X}^2_i E_{\hat{v}_i} \bar{X}^2_j$ for every $i \in \{2,\ldots,n\}$ and every $j \in \{1,2,\ldots,n\}$.
				
				\textbf{Case B:} Suppose $X^2_1=W^2$. Consider the allocation $\bar{X}^2_N$ as defined in the previous case (Case A).

				Therefore, in both cases, Case A and Case B, we have an allocation $\bar{X}^2_N$ such that
				\begin{enumerate}
					\item [(i)] $\bar{X}^2_1=W^2$,
					\item [(ii)] $|\bar{X}^2_i|\in \{\max(X^1_N)-1,\max(X^1_N)\}$ for all $i \in \{2,\ldots,n\}$, and
					\item [(iii)] $\bar{X}^2_i E_{\hat{v}_i} \bar{X}^2_j$ for every $i \in \{2,\ldots,n\}$ and every $j \in \{1,2,\ldots,n\}$.
				\end{enumerate}
				
				Therefore, either $\bar{X}^2_N$ is EFX at $\hat{v}_N$ or $\bar{X}^2_1 E_{\hat{v}_1} \bar{X}^2_i$ does not hold for some $i \in \{2,\ldots,n\}$. If $\bar{X}^2_N$ is EFX at $\hat{v}_N$, then the proof is complete. Suppose $\bar{X}^2_N$ is \textit{not} EFX at $\hat{v}_N$. Therefore, it must be the case that $\bar{X}^2_1 E_{\hat{v}_1} \bar{X}^2_i$ does not hold for some $i \in \{2,\ldots,n\}$.
				
				Recall that $|X^1_1|< |X^1_i|$ for every $i \in \{2,\ldots,n\}$. Therefore, it must be the case that $|\bar{X}^2_1| \leq \max(X^1_N)-1$ because $|\bar{X}^2_1|=|\bar{X}^1_1|+1$. Since $\bar{X}^2_1 E_{\hat{v}_1} \bar{X}^2_i$ does not hold for some $i \in \{2,\ldots,n\}$, it must be the case that $|\bar{X}^2_1| < \max(X^1_N)-1$. To see this, suppose $|\bar{X}^2_1| = \max(X^1_N)-1$. Then $\max(X^1_N)=3$. Suppose $\bar{X}^2_1 E_{\hat{v}_1} \bar{X}^2_j$ does not hold, where $j \in \{2,\ldots,n\}$. Then, $\hat{v}_1(\bar{X}^2_j \setminus \{x_j\})> \hat{v}_1(\bar{X}^2_1)$ for some $x_j \in \bar{X}^2_j$. Hence, there exists $Z \subseteq E \setminus W^2$ with $|Z|=2=\max(X^1_N)-1$ such that $\hat{v}_1(Z)>\hat{v}_1(W^2)>\hat{v}_1(X^1_1)$, which is a contradiction to the definition of $W^2$ because $Z \in G^1$ and $\hat{v}_1(Z)>\hat{v}_1(W^2)$.\footnote{Notice that such $Z$ exists because for the case where $|\bar{X}^2_j|=2=\max(X^1_N)-1$, we can choose $Z=\bar{X}^2_j$ and for the case where $|\bar{X}^2_j|=3=\max(X^1_N)$, we can choose $Z=\bar{X}^2_j \setminus \{x_j\}$.} Therefore, $|\bar{X}^2_1| < \max(X^1_N)-1$.

				We slightly abuse the notation and use $X^2_N$ to denote $\bar{X}^2_N$. Using similar arguments as before, we conclude that there exists $Z \subseteq E \setminus X^2_1$ with $|Z|=\max(X^2_N)-1$ such that $v_1(Z) > v_1(X^2_1)$. Then, we construct non-empty sets $W^3$ and $Y^3$ in similar fashion where $|W^3|=3$. We can write $m-2=(n-1)\ell+r$ for some $\ell \in \mathbb{N}$ and $0\leq r <n-1$.\footnote{We again slightly abuse the notation here as we are using $\ell$ and $r$ here that might be different than the previously written $\ell$ and $r$.} Following similar arguments as before and considering Case A and Case B as before, either EFX exists at $\hat{v}_N$ or we get hold of $X^3_N$ (defined in a similar way as before) such that 
				\begin{enumerate}
					\item [(i)] $X^3_1=W^3$,
					\item [(ii)] $|X^3_i|\in \{\max(X^2_N)-1,\max(X^2_N)\}$ for all $i \in \{2,\ldots,n\}$, and
					\item [(iii)] $X^3_i E_{\hat{v}_i} X^3_j$ for every $i \in \{2,\ldots,n\}$ and every $j \in \{1,2,\ldots,n\}$.
				\end{enumerate} 
				
				Therefore, either $X^3_N$ is EFX at $\hat{v}_N$ or $X^3_1 E_{\hat{v}_1} X^3_i$ does not hold for some $i \in \{2,\ldots,n\}$. If $X^3_N$ is EFX at $\hat{v}_N$, then the proof is complete. Suppose $X^3_N$ is \textit{not} EFX at $\hat{v}_N$. Therefore, it must be the case that $X^3_1 E_{\hat{v}_1} X^3_i$ does not hold for some $i \in \{2,\ldots,n\}$. Using similar arguments as before, we can establish that $|X^3_1| < \max(X^2_N)-1$.
				
				We continue this process. Since $E$ and $N$ is finite, either EFX exists at $\hat{v}_N$ or there exists $c \in \mathbb{N}$, non-empty sets $W^c$ and $Y^c$ where $|W^c|=c$, together with an allocation $X^c_N$ (defined in a similar way as before) such that 
				\begin{enumerate}
					\item[(i)] $c\leq |X^c_i|\leq c+1$ for each $i \in N$,
					\item [(ii)] $X^c_1=W^c$,
					\item [(iii)] $|X^c_i|\in \{\max(X^{c-1}_N)-1,\max(X^{c-1}_N)\}$ for all $i \in \{2,\ldots,n\}$, and
					\item [(iv)] $X^c_i E_{\hat{v}_i} X^c_j$ for every $i \in \{2,\ldots,n\}$ and every $j \in \{1,2,\ldots,n\}$.
				\end{enumerate}

				Once again, either $X^c_N$ is EFX at $\hat{v}_N$ or $X^c_1 E_{\hat{v}_1} X^c_i$ does not hold for some $i \in \{2,\ldots,n\}$. We claim that $X^c_N$ must be EFX at $\hat{v}_N$. To see this, note that $X^{c-1}_1 E_{\hat{v}_1} X^{c-1}_i$ does not hold for some $i \in \{2,\ldots,n\}$. If this is not the case, then $X^{c-1}_N$ would be EFX at $\hat{v}_N$ and the process would have terminated at $c-1$. Following similar arguments as before, it must be the case that 
				\begin{enumerate}
					\item [(i)] $X^{c-1}_1=W^{c-1}$ where $|W^{c-1}|=c-1$,
					\item[(ii)] $|X^{c-1}_i|\in \{\max(X^{c-2}_N)-1,\max(X^{c-2}_N)\}$ for all $i \in \{2,\ldots,n\}$, and 
					\item[(iii)] $|X^{c-1}_1| < \max(X^{c-2}_N)-1$.
				\end{enumerate}
				
				Notice that by definition, $\hat{v}_1(X^1_1)<\hat{v}_1(X^2_1)<\ldots<\hat{v}_1(X^c_1)$ and $|X^1_1|<|X^2_1|<\ldots<|X^c_1|$ where $|X^j_1|=|W^j|=j$ for every $j \in \{1,\ldots,c\}$. Also note that $\max(X^{c}_N) \leq \max(X^{c-1}_N)\leq \ldots \leq \max(X^{2}_N) \leq \max(X^{1}_N)$ and whenever $\max(X^{j+1}_N) < \max(X^{j}_N)$ for some $j \in \{1,\ldots,c-1\}$, it must be the case that $\max(X^{j+1}_N)+1=\max(X^{j}_N)$. Since, $c\leq |X^c_i|\leq c+1$ for each $i \in N$ and $|X^c_i|\in \{\max(X^{c-1}_N)-1,\max(X^{c-1}_N)\}$ for all $i \in \{2,\ldots,n\}$, it must be the case that $\max(X^{c-1}_N)=c+1$ because if $\max(X^{c-1}_N)=c$, then $|X^{c-1}_i|\in \{\max(X^{c-2}_N)-1,\max(X^{c-2}_N)\}$ for all $i \in \{2,\ldots,n\}$ implies that $\max(X^{c-2}_N)=c$ which would in turn imply that $|X^{c-1}_1|=\max(X^{c-2}_N)-1$, a contradiction to the fact that $|X^{c-1}_1| < \max(X^{c-2}_N)-1$. Therefore, it must be the case that $\max(X^{c-1}_N)=c+1$.
				
				Now, we show that $X^c_1 E_{\hat{v}_1} X^c_i$ holds for all $i \in \{2,\ldots,n\}$, thereby establishing that $X^c_N$ is EFX at $\hat{v}_N$, thus completing the proof. Suppose $X^c_1 E_{\hat{v}_1} X^c_j$ does not hold, where $j \in \{2,\ldots,n\}$. Then, $\hat{v}_1(X^c_j \setminus \{x_j\})> \hat{v}_1(X^c_1)$ for some $x_j \in X^c_j$. Hence, there exists $Z \subseteq E \setminus W^c$ with $|Z|=c=\max(X^{c-1}_N)-1$ such that $\hat{v}_1(Z)>\hat{v}_1(W^c)>\hat{v}_1(X^{c-1}_1=W^{c-1})$, which is a contradiction to the definition of $W^c$ because $Z \in G^{c-1}$ and $\hat{v}_1(Z)>\hat{v}_1(W^c)$ where $W^{c}=\arg\hspace{-2mm}\max_{S \in G^{c-1}}\hat{v}_1(S)$ and
				$$G^{c-1}=\big \{S \subseteq E : |S|=c \mbox{, and } \exists T \subseteq E \setminus S \mbox{ with } |T|=\max(X^{c-1}_N)-1 \mbox{ and } \hat{v}_1(T)>\hat{v}_1(X^{c-1}_1)\big \}.$$.
				
				Notice that such $Z$ exists because for the case where $|X^c_j|=c=\max(X^{c-1}_N)-1$, we can choose $Z=X^c_j$ and for the case where $|X^c_j|=c+1=\max(X^{c-1}_N)$, we can choose $Z=X^c_j \setminus \{x_j\}$.

				Therefore, it must be the case that $X^c_1 E_{\hat{v}_1} X^c_i$ holds for all $i \in \{2,\ldots,n\}$. Hence, $X^c_N$ is EFX at $\hat{v}_N$. This completes the proof of the theorem.
			\end{proof}

		}

	\end{appendices}

	\bibliographystyle{plainnat}
	\setcitestyle{numbers}
	\bibliography{mybib}

\end{document}